\documentclass[11pt]{article}

\usepackage{graphicx}
\usepackage{amsmath}
\usepackage{amssymb}
\usepackage{amsthm}
\usepackage{caption}
\usepackage{color}
\usepackage{algorithm2e}
\usepackage{enumerate}
\usepackage{multicol}
\usepackage[margin=0.75in]{geometry}
\theoremstyle{definition}

\newtheorem{thm}{Theorem}[section]

\newtheorem{prop}[thm]{Proposition}

\newenvironment{dfn}{\refstepcounter{thm} \bigskip \par \noindent
    {\bf Definition \thethm}\ }{\bigskip \par}
\newenvironment{exa}{\refstepcounter{thm} \bigskip \par \noindent
        {\bf Example \thethm}\ }{\bigskip \par}
\newenvironment{pf}{\medskip \par \noindent {\it Proof.}\ }{\hfill
        \P \bigskip \par}

\title{Incentive-based integration of useful work into blockchains\\(Working draft)}
\date{June 2018}
\author{David Amar\footnote{For comments, questions, and suggestions please contact ddam.am AT gmail.com} \and Lior Zilpa}

\begin{document} 

\maketitle

\section*{Abstract}
Blockchains have recently gained popularity thanks to their ability to record \''digital truth\''. They are designed to keep persistence, security, and avoid attacks which is useful for many applications. However, they are still problematic in their energy consumption, governance, and scalability Current solutions either require vast computing power via Proof-of-Work (PoW) or cannot directly utilize computing power as a resource in virtual mining. 

Here, we propose incentive-based protocols that use competitions to integrate computing power into blockchains. We introduce Proof-of-Accumulated-Work (PoAW): miners compete in costumer-submitted jobs, accumulate recorded work whenever they are successful, and, over time, are remunerated. The underlying competition replaces the standard hash puzzle-based competitions of PoW. A competition is managed by a dynamically-created small masternode network (dTMN) of invested miners. dTMNs allow for scalability as we do not need the entire network to manage the competition. Using careful design on incentives, our system preserves security, avoids attacks, and offers new markets to miners. When there are no costumers the system converges into a standard protocol. 

Our proposed solution improves the way by which the blockchain infrastructure works and makes use of its computing power. We also discuss how the protocol can be used by fields that require solving difficult optimization problems, such as Artificial Intelligence and Pattern Recognition in Big Data. \\

\textbf{Keywords}. Blockchain, computing, algorithms, useful work, artificial intelligence, competitions.

\begin{multicols}{2}

\section{Introduction}

In this section we provide a thorough review about blockchains and the difficulty in utilizing their computing power. Readers familiar with basic terminology can skip this section.

\subsection{General Background}

Bitcoin, presented first in 2008, was the first  functional decentralized digital cash system to achieve international popularity \cite{Nakamoto2008}. It is based on an infrastructure called \textit{blockchain}: an immutable ledger of transactions made among public keys of users. As a result, Bitcoin and other blockchain based cash systems are also called cryptocurrencies. In recent years, cryptocurrencies and blockchains have gained popularity but were also criticized for being used in a speculative nature, and for lacking substance \cite{Coindesk}. Nevertheless, the interest in the blockchain technology itself is high as it can be utilized for decentralized self-governing protocols. The applicability of such systems is being investigated in other domains such as smart contracts, credit systems, artificial intelligence, and others \cite{McConaghy2017}. 

A blockchain is a \textit{persistent decentralized} ledger of transactions. Persistence, in this context, means that: (1) there is an underlying data structure that keeps an ordered collection of records, typically transactions, and (2) past transactions cannot be modified (i.e., immutability). Thus, this data structure essentially represents a recorded digital truth \cite{Godsiff2015}. The term blockchain is used because the data are kept in a linked list of blocks of information that can be extended only by adding a new single block as the head of the chain. Decentralization in our context means that there is no single entity that keeps the information and to which all trust is given. Alternatively, there is a network of peers that each holds a copy of the data. This is accompanied with a protocol that lists rules on how a new block is added. 

Blockchains can be public or kept within an organization. Regardless of the paradigm, they still have a network of peers, also called \textit{miners}, and possible users. To keep the properties above, data must be accessible to all network members and users. Users can get multiple copies of the ledger and use the protocol to decide which one is correct. The accepted majority version is called the consensus. In addition, blocks are added in a controlled rate to guarantee that there is enough time for the network to reach a consensus when the data structure changes. 

Miners both keep a copy of the blockchain and can propose new blocks. They have a clear incentive to have their newly proposed blocks accepted to become a part the blockchain (i.e., in the copy of every other miner): they are entitled for a payment. As a result, miners \textit{compete} for the right to add new blocks. This competition seems to be in contradiction to the need to regulate the block addition rate, which is essential to keep decentralization. This gap is solved by introducing the concept of proofs: for the network to consider a new block, the miner must prove to all other peers that a certain resource was spent in the process. Using proofs achieves another important goal. Since the miners compete amongst themselves, the proof system makes it expensive to achieve exclusive ownership on block addition. This property keeps the integrity of the data by establishing trust among the miners. 

Public blockchains must be resistant to attacks in which an adversary tries to tamper with the data by either introducing false transactions or falsifying past data, which can be used to carry double spending. One source of susceptibility for such attacks are cases in which there is a miner that holds 51\% or more of the resources on which proofs rely. If such a situation is used for falsifying data it is called a \textit{51\% attack}. However, note that even if such an attack succeeds it will be easily visible by the users and other miners because the data are public. Thus, if there is a large set of honest miners that abide to the protocol, they will reject the false blocks and the network will split into two versions. A split is both visible and is likely to reduce the value of the entire system, including for the adversary.

Popular blockchain systems today are based on Proof-of-Work (PoW): proofs that guarantee that a certain amount of computing work was performed. This is done using a well-known cryptographic tool called hash-based computational puzzles, for more details see \textbf{Section 1.3}. PoW protocols are designed such that the expected number of blocks that miners sign is proportional to their relative weight compared to the total computing power of the network. As a result, 51\% attacks are achievable only by a miner whose total computing power is larger than 50\% of the combined power of the network. An alternative to PoW is called virtual mining on which we focus below.

\subsection{Virtual Mining}

The basic idea of virtual mining is that money itself can be treated as a form of PoW. In other words, the act of acquiring coins proves that peers had invested from their own resources to become a part of the system. Therefore, we can use money, either owned or deposited, as the underlying proof system of the blockchain. Moreover, deposited amounts can be used as collateral, thereby promoting trust. Such systems are often called Proof-of-Stake (PoS) systems, but other related names such as Proof-of-Deposit (PoD) are also used. Naturally, as in PoW, the system will give the right to sign blocks in a way that is proportional to the amount of stakes each miner presents. As a result, PoS has a security advantages: it reduces 51\% attacks. Such attackers need to get coins first at large amounts. This both require substantial funds in other coin (crypto or not) and raises the value of the coin. Together it makes it less likely that such attacker will gain from attacking the system.

It is important to note that virtual mining has some issues that are still under research. The most famous example is multiforging that results from Nothing-at-Stake (N@S): miners in PoS have nothing to lose by approving different histories \cite{Chepurnoy2015}. PoW avoids that by having the solved puzzles dependent on the information of the suggested block. Moreover, it is expensive and not advantageous to work in different chains. In contrast, in PoS we cannot force miners to converge into a single chain because miners do not lose anything by multi-branching. On the other hand, virtual mining involves stacking a large amount of coins over time to achieve reasonable weight in the system. Thus, such miners both lose the liquidity of their resources and are highly invested in the success of the network. This somewhat contradicts N@S in that a network with multiple simulated chains will potentially get bad publicity and lose some of its value. We list below some extant virtual mining solutions.

Peercoin (PPCoin) by King and Nadal, 2012 \cite{King2012} is a PoW-PoS hybrid system. There are two block types: PoW-based and PoS-based. PoS blocks have a special transaction in which the signer pays himself (thereby nullifying the age of his coins). This transaction is called coinstake and its first input is called the kernel. The kernel is a solution to an easy hash puzzle. The hash puzzle difficulty in the kernels is inversely proportional to the amount of stakes that the miner burns. The difficulty formula was designed to give preference to stakes over pure hash solutions. The advantage of this approach is that utilizing coin-age within the protocol means that unlike other PoS systems coins can be staked once, which makes the system more decentralized. On the other hand, this system has a centralized mechanism called checkpoints in which a group of trusted miners time-stamp a version of the blockchain. This idea was taken from one of the earliest modifications to Bitcoin, and is used to prevent N@S issues. As a side note, this implies that Bitcoin and Peercoin are not fully decentralized systems.

DASH by Duffield and Diaz \cite{Duffield} is a method that was designed as a PoW system that overcomes some of the main problems in Bitcoin including: scalability, privacy issues, and the decrease in the number of service providing nodes. The method introduced the notion of a secondary network of masternodes: full nodes that have incentives to provide services to users and not only deal with mining. To become a masternode the miner has to store 1000 DASH coins. In return to providing special services to the network masternodes are paid roughly 45\% of the total block reward. Thus, the masternodes act as a quorum of trusted nodes that can perform special transactions such as mixing, which is a technique that allows anonymizing client's history, and instant sending of funds. DASH is an elegant PoW and Proof-of-Service (PoSe) combination that promotes decentralization in a secured way. However, the idea of PoSe-based secondary master network is a general idea that can be exploited for other tasks.

Hybrid methods: Decred \cite{Jepson2015} – this system was proposed by the developers of btcsuite. This is a PoW/PoS hybrid system that aims to solve N@S by having PoS miners acting as a second authentication mechanism for new blocks. Thus, PoS cannot be used for creating multiple chains because it only approves past blocks. Decred also strengthen governance because the external PoS miners act as another layer that votes on the integrity of the data.

Ethereum's Casper – this is a complex protocol that has not been materialized at this point but shares some conceptual similarity with Decred. Ethereum \cite{Buterin2015} set a goal is to implement a PoS system that will be integrated and finally replace PoW. The proposed PoS has two features that make Casper different from other suggestions. First, unlike Peercoin, to become a miner there is a minimal sum that must be staked.  Second, the network punishes an attacker by consuming the coins of his stakes. Thus, unlike the other PoS alternatives it directly reduces the incentives of an attacker. Moreover since creation of multiple chains is easily detectable, the network can easily punish miners that attempt such attacks.

\subsection{Hash puzzles}

PoW uses computational puzzles that are based on hash functions, which makes them practically useless for other applications (other than those used in cryptography and blockchains). 

\dfn{
A hash function $H$ maps input strings to output strings and have the following five properties \cite{Narayanan2016}:
\begin{enumerate}
	\item \textit{Any input, fixed output}: $H$ takes an input of any size and outputs a string of a fixed size $n$. That is, $H:\{0,1\}^* \rightarrow \{0,1\}^n$. 
	\item \textit{Collision resistance}: it is computationally infeasible to find two inputs $x$ and $y$ such that $x \ne y$ and $H(x)=H(y)$. 
	\item \textit{Hiding}: It is infeasible to learn what $x$ is from knowing $H(r||x)$ where $||$ denotes concatenation and $r$ is a randomly selected value from a distribution $F$ whose minimal entropy is high. 
	\item \textit{Puzzle friendliness}: there is no algorithm with an average case complexity better than $o(2^n )$ to find $x$ given $H(k||x)$, where $k$ is a randomly selected value from a distribution $F$ whose minimal entropy is high.
	\item Cheap validation: for any input $x$ of size $m$, $H(x)$ is computable in $O(m)$.
\end{enumerate}
}

Hash puzzles are only good for showing that the miner performed many computations and cannot be used for practical computational tasks. This major drawback can prevent using PoW in very large systems as the energy consumption can be too large. On the other hand, the combined computational power that results from such networks is a valuable resource that can be utilized as another resource on which the blockchain is built.

\subsection{Useful work challenges}

One inevitable question is why is it not straightforward to utilize the blockchain network for solving real computational problems? This question depends on the underlying blockchain context. For PoW, can hash puzzles be replaced with useful computational work? For PoS, can we implement an efficient contract-based system? 

First, useful computational work is not puzzle friendly nor easy to track. Puzzle friendliness, on the other hand, requires having a tight lower bound on the running time for each puzzle instance. This is a very strong constraint that is not expected to be satisfied in an arbitrary computational problem. Even if we have theoretical results for a specific problem type, it usually does not guarantee that the theoretical complexity will be realized in each instance. Moreover, when solving optimization problems we often do not know if a given solution is the best one (or close enough to the best one). For example, for NP-Complete (NPC) problems, we may be able to validate that a given solution is correct but we do not know that it is necessarily the best one. More formally, for a NPC problem $M$ with a polynomial validation algorithm $F_M(s)$ that efficiently computes the score of a candidate solution $s$, we can only validate $F_M(s)=x$, where $x\in{0,1}$ for a decision problem, and a score for an optimization problem. However, we often cannot claim that there is no other solution with a better score, unless NP=coNP. Non-convex optimization problems have a similar problem: we may know that we had reached a local optimum, but we often do not know if it is a global optimum or not. Another issue is that for many applications the underlying problems depend on high volume data, which is out of the scope of the block sizes of almost all current blockchains. Thus, a complete solution must utilize a storage component. 

Second, a simple contract-based system will not directly utilized the incoming resources to strengthen the blockchain, nor will it exploit more than a single market for the miners. While implementing a computing resource or a hardware leasing application on top of a blockchain seems practical, a miner that provides hardware acts as a contractor and has no responsibilities nor incentives to maintain the network. Naturally, an integrated system in which miners are paid both for useful work and for being peers in the network will be more attractive.

Third, hybrid systems and systems that aim to deviate hash energy to contact-based useful work are directly open to a precarious vulnerability to adaptive 51\% attacks: powerful miners can monitor the hash difficulty and wait for time points in which it is feasible for them to carry a 51\% attack. This sensitivity stems from inflated variance in hash puzzle difficulty due to oscillations in the percentage of energy devoted to puzzles. 
This point may seem subtle. However, it creates a major problem for contract-based methods as they may seem to take the blockchain infrastructure for granted and in the same time open it to attacks. Moreover, this problem may also affect PoS systems that are based on pricing the stakes. As we shall explain below, we solve this issue by introducing a new system based on accumulating useful work.

Fourth, a protocol that rewards computational problem solvers with preference for signing blocks is exposed to new attacks:

\dfn{An $O(1)$ attack of a protocol $P$ is enabled if $P$ either allows or enables a state with a positive incentive for an adversary to: (1) post problems for which he knows the optimal answer, and (2) immediately propose solutions to these problems. In such cases we say that $P$ is $O(1)$-sensitive. The adversary's positive incentive can be direct (e.g., $P$ pays for solving tasks) or indirect such as by gaining preference in adding new blocks.}

\dfn{A suboptimal solution adversary (SSA) for a problem $M$ in a protocol $P$ is a miner that knowingly posts suboptimal or intermediary solutions for $M$. The protocol $P$ is SSA-enabling if there is no negative result for posting such solutions.}

Finally, hash functions cannot be simply replaced with other functions that are useful for real-world applications without tampering security. Intuitively, real-world applications are based on functions that preserve information: similar inputs are likely to result in similar answers, and the output usually gives some information about the output. Thus, as a general rule of thumb, useful functions unlike hash functions are not collision resistant nor hiding. For an example from distances in graphs, see the \textbf{Appendix}.

\subsection{Extant systems}

Previous work addressed useful work mainly by trying to replace hash functions with other computations. For example, Oliver et al., 2017 \cite{Oliver2017} proposed solving NP complete problems (NPC). Note, however, that much like the graph distance problem above, such problems do not satisfy collision resistance and hiding, and only partially satisfy puzzle friendliness and cheap validation. For validation the problem is that while NPC solutions can be validated to be true once cannot guarantee that a negative proposition (i.e., there are no cliques of size k in the graph G) unless NP=coNP. As a result, Oliver et al., 2017 \cite{Oliver2017} proposed that a new block can be signed with a solution to a problem P only if the solution improves upon past solutions.  Surprisingly, using the $O(1)$ and $SSA$ described above the blockchain system of \cite{Oliver2017} can be easily damaged such that: (1) the block addition rate cannot be regulated, and (2) an adversary can take over many blocks without performing computations. 

Another problem of the approach above is that difficulty of instances of NPC problems is hard to determine in advance (even though rough bounds can be computed), and thus even without attacks the control of block addition may be jeopardized. Several previous papers have presented ways to preserve hash properties by substantially limiting the set of practical problems that can be solved. Ball and colleagues \cite{Ball2017} presented a proof system based on solving the orthogonal vectors problem, whereas \cite{King2013} proposed primecoin:  a system based on looking for Cunningham numbers.

\section{Our new PoAW protocol}

\subsection{Outline}

We start by introducing the idea of Proof-of-Accumulated-Work (PoAW): a system that accumulates past useful work in a way that is secure thanks to protocol-enforced constraints and rewards. Second, we introduce dynamic Task Masternode Networks (dTMNs): dynamically created subnetworks of invested miners that manage user-submitted tasks.

To achieve our goals we integrate our new concepts with: (1) virtual mining, (2) replacing competitions for solving hashes with competitions over computational tasks, and (3) storage blockchains. Virtual mining (with or without using hybrid PoW within) is used to keep the main blockchain and acts as the glue between the different concepts. Public competitions are the places in which computational miners can do useful work and reach an additional market other than the financial system of the blockchain itself. The storage chain keeps the information about the computational tasks: the input data and the solutions submitted by the competing nodes throughout each competition. Miners of the distributed storage can constitute the dTMN, which also validates suggested solutions and get rewards. 

We also avoid the need to precisely evaluate the difficulty of the users' computational tasks. Alternatively, we let users offer the tasks to the system and their payment. Thus, we let the market regulate the value of each optimization problem that is submitted.

The resulting protocol achieves three important goals. First, the network keeps the cryptographic power of the blockchain components intact. Second, most of the computational power of the system can be utilized to solve optimization problems. Finally, miners earn from both serving the blockchain and solving computational tasks. Moreover, we prove that joining our network as a computational node is more profitable than either renting hardware in a contract-based system or standard PoW mining.

\subsection{The virtual mining system}

\subsubsection{Virtual stakes}

We use virtual mining to obtain two goals. First, as in standard virtual mining, miners can use their stakes to sign or approve new blocks, which has beneficial results for a decentralization and governance \cite{Jepson2015}. Second, it acts as a device for implementing PoAW within the protocol. Miners that solve computational tasks get \textit{virtual stakes} (vstakes) once a competition is sealed. Thus, when a computational task is sealed, there is a special transaction that creates a certain amount of locked virtual money that the miner can use as stakes only. These are not coins that the user can send or spend: these are rather specialized tokens that the miner gets from the system in return for solving tasks. These amounts can be accumulated, which means that by solving computational tasks miners transform their useful work into system-approved certificates. 

\textit{Vstakes} are a general way to accumulate work. They provide a way to later remunerate miners for their work, which already earned from solving fees. Moreover, they can be utilized within any PoS or hybrid PoW/PoS system as long as it allows recording and burning virtual stakes. Specifically, in our current system we propose using virtual stakes to approving blocks within the PoS system of Decred (see \textbf{Figure 1} for illustration).

\begin{figure*}
  \centering
  \includegraphics[width=115mm,height=55mm]{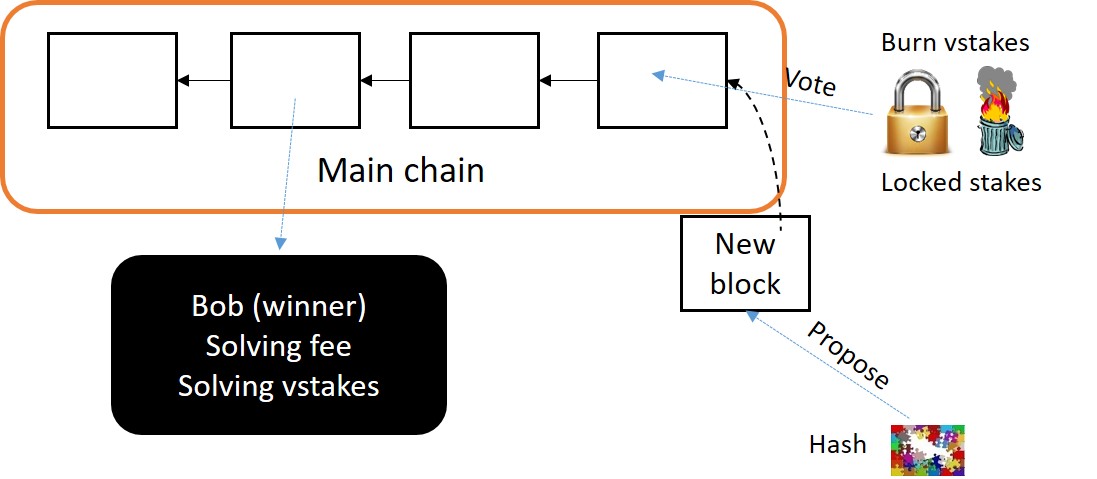}
  \caption{Computational miner Bob wins a competition for a task and gets virtual stakes. To later sign a new block Bob can burn previously earned virtual stakes, used locked real money stakes (as in the standard PoS protocol), or present hash puzzle solutions.}
  \label{Figure1}
\end{figure*}

\subsubsection{Decred augmentation}

Decred \cite{Jepson2015}, which shares conceptual similarities to both DASH \cite{Duffield} and Peercoin \cite{King2012}, is based on a lottery system in which tickets are purchased and then deposited for a wait period before they can be spent. Selection of tickets from a pool is done using a pseudorandom algorithm that is based on information in the head of the new block. This allows (pseudo) random selection of tickets from the pool. Each ticket gives the PoS miner a single vote, but ticket prices are determined by the protocol automatically such that the total number of tickets is maintained at a desired limit. This process is the Decred PoS equivalent of a standard hash difficulty determination algorithm. For a new block to be accepted into the blockchain five votes must approve it. The PoS miners receive both their ticket’s worth and additional payment. Note that there is an approximately 5\% chance that a ticket will end up deleted without producing additional reward.

A ticket’s lifetime goes through the following stages. It is created when a miner pays the \textit{ticket price} $+$ \textit{ticket fee}. A block can create up to $20$ new tickets. Tickets are mined into a block such that those with higher fees get preference. Once accepted into the blockchain, a ticket must wait $256$ blocks before it enters into the ticket pool. The ticket pool size is kept at $40960$. If a ticket is not selected after $142$ days it expires and the miner is paid back the ticket price. After a ticket is consumed, it enters into another $256$ blocks period of wait after which its associated funds are released. 

Voting is done by transforming tickets into vote transactions. A vote represents the ticket validation of the last seen block. The new block must contain a majority of the votes to be added into the blockchain (which implies that votes can be missed). Only when the previous block is validated by the votes in the new block the block’s funds in the UTXO set (set of remnants funds and rewards to the miner) are allowed. The process presents the validation result of the last block in a single bit that summarizes the result. Also, note that PoS miners must remain constantly active as they may otherwise miss their voting turn. This property of Decred will be discussed again in the next section. 

Decred is similar in spirit to DASH in that they both establish a masternode network that takes over a desired task. In both, becoming a member of this network costs either in tickets in Decred or in depositing $1000$ coins in DASH. Note that DASH is not a PoS system per se in that there is a single deposit and amounts are not staked. Nevertheless, the idea of substantial initial deposit can be easily extended into the Decred protocol. Also, note that the main goal of Decred is to strengthen decentralization and open governance while promoting community input. These needs stem from past weaknesses observed in Botcoin and past criticism of the PoS idea. On the other hand, Decred was not developed to deal with the energy consumption problem and the non-usefulness of PoW.  

In our augmentation of Decred vstakes are given to the winners of a computational competition at its last transaction, called $Seal_{CT}$. Let $fee_{solve}$ be the amount that the client wishes to pay for a task $CT$ (a formal definition of a task and its associated transactions will be give below). We define a protocol parameter called $P_{vstake}$. The $Seal_{CT}$ transaction contains a special minting command that gives the following amount of vstakes to each winner: $$\frac{(P_{vstake}*fee_{solve})}{N_W}$$ where $N_W$ is the number of miners that won the competition.

\subsubsection{The PoS higher profit invariant}

Virtual stakes can be used only for purchasing tickets. Moreover, we define another constraint that is essential for preventing $O(1)$ attacks. Let $x$ be the amount in coins of virtual stakes that are used to purchase a ticket. Our \textit{PoS higher profit invariant} (simply called PoS invariant) determines that $x$ cannot yield extra profit from the system. 

Formally, let $y>x$ be the price of a ticket and let $r>1$ is the expected profit factor of the PoS mining. For example, in a system that is designed to produce an expected 10\% profit on PoS spent money $r=1.1$. In our system only $y-x$ is entitled to receive $r$ where $x$ is entitled to exactly $1$. That is, the PoS miner will receive from the system an expected reward of $x+r(y-x)$. Another way to interpret this constraint is that PoW miners are not asked by the protocol to share their profits with the PoS rights that result from virtual stakes.

In our protocol each new block also mints new coins. Note that these new coins are not entitled to virtual stakes. Thus, PoW miners are paid once for their work. However, they do not get all minted coins as in Decred, some of these sums go to the PoS miners as a reward.

\subsubsection{The robustness of the system}

By accumulating useful work within virtual mining we obtain a way to ensure the strength of the system even if a hybrid PoS/PoW like Decred is used. The accumulation and scoring of past work using vstakes means that adversaries that try to reach considerable resource weight now compete with both miners that solve hash puzzles,  miners that solve computational tasks, and miners that do pure PoS mining. Moreover, the integration with Decred means that getting double spending approved requires reaching 51\% of the combination of all three resources above.

First, our competition system explained in the next section prevents $O(1)$ attacks and reduces the incentive to submit suboptimal solutions. Second, note that our system further avoids N@S (i.e., on top of the specific system that is used, e.g., DASH or Decred) by directly mixing stakes with useful work vstakes that are used for proof of burn: miners can spend their hard earned vstakes once and then they are deleted. This is equivalent to paying in actual cash for the right to sign blocks. Thus, miners are not likely to spend their vstakes in different chains as some may be lost from the consensus.

\subsection{Computational competitions}

\subsubsection{Preliminaries and notations}

\textit{Tasks}. A computational task $CT=<Type,F_D,D,C,S>$,  is a five-tuple where $Type$ is a bit specifying if this is a maximization or a minimization problem, $F_D$ is the scoring function we wish to optimize that depends on a dataset $D$, which can be null if $F_D$ is self-contained. $C$ is a set of constraints, and $S$ is a set of additional search space parameters on top of the ones implied by the definition of $F_D$. For each computational task $CT$ we define its slim version $CT^{s}=<Type,F_D,H(D),C,S>$ which is exactly as  $CT$ except for the data element: we keep the hash result of the data instead of $D$ itself. Thus, $CT^s$ requires much less space as compared to $CT$. 

\exa{An instance of an Integer Linear Programming (ILP) problem has a linear optimization function we wish to maximize $F=c^T x$, $D=\emptyset$. $C$ is specified using the following constraints:
$$Ax \le b$$
$$x\in Z^n$$
 ILP is NPC even if $x\in\{0,1\}^n$. 
A generalized version of ILP that allows both discrete and continuous variables is called Mixed Integer Linear Programming (MILP).
}

\exa{\textbf{(GD)} Let $G$ be a continuous and differentiable function in $R^n$. Gradient descent (GD) algorithms define an optimization problem for minimizing G that is based on computing the steepest descent. Let $\nabla G$ be the gradient function of G. Given a point $x\in R^n$ in the search space we update it by computing:
$x'=x- \epsilon \nabla G(x)$
where $\epsilon$ is the learning rate of the algorithm. The algorithm converges to a local minimum when the difference between $x$ and $x'$ is negligible. A simple way to reformulate the problem using our notation is to let miners search for local optimum points defined as points in which $x'-x \le \epsilon^*$, which means $\epsilon \nabla G(x)\ge \epsilon^*$. Thus, we wish to maximize $F(x)= \epsilon \nabla G(x)$. We can further define in $S$ the allowed approximation accuracy on $x$, e.g., do not use resolution greater than $10^{-3}$. Of course, this formulation, while being legal, distorts the original goal of minimizing $G$. Clearly we can have $F(x)=G(x)$ and let miners choose their own algorithm, but the goal of this example was to illustrate the applicability for GD instances. 
}\\

\textit{Transactions}. We define a set of competition-related transactions. $Publish_{CT}$ is a bid offer by the client to the storage. $Stored_{CT}$ is a transaction that marks the acceptance of $CT$ and its associated dataset into the storage. It also marks the acceptance of $CT$ into the system as a new competition. $Solve_{CT}$ is a transaction that contains an offered solution. $Validate_{CT}$ is a transaction in which the \textit{dynamic Task Masternode Network (dTMN)}, which is a group of storage peers that govern data storage and validation, is asked to validate and choose the winner, which is finally marked using a transaction called $Seal_{CT}$. 

As common in blockchains, the transactions above may offer some fees to promote their execution or addition to the ledger. Generally, our transactions have two mechanisms by which service fees are paid. First, we have regular fees that are sums that party A sends directly to party B. These are generally denoted as $Fee_{tr}$. Second, we have \textit{promised fees (PFs)}, which are sums paid later than the block if a certain criteria was met.

Formally, a promised fee $PF=<o,p,b>$ is a contract stating that if a payment $o$ is carried then a certain fraction $p\in [0,1]$ is sent after at least $b$ blocks. 
Formally, all competition related transactions are tuples and each transaction has a unique identifier which can be denoted as $ID(Tr)$ of a transaction $Tr$ (of any type). We define several transactions below. For ease of notation we omit the transaction ID and its $Fee_{tr}$.

\begin{itemize}

\item The first transaction is called $Publish_{CT}$ and it gives all the needed information from the client. $Publish_{CT}= < CT^s,Fee_{sub},Fee_{solve},PF_{solve},PubID_c>$, where $CT^s$ is the slim version of $CT$, $Fee_{sub}>0$ is a storage submission fee. $Fee_{solve}$ is the total offered solving fee, $PF_{solve}$ is a a data structure that details the promised fees taken from $Fee_{solve}$ as reward to the infrastructure maintaining miners. Note that $PF_{solve}$ specifies amounts that are paid only if the competition successfully outputs a result.This issue is further discussed in \textbf{Section 3}.
\item $Stored_{CT}= < CT^s,ID_{publish},PubID_c,\\ PubID_{masternode}>$, where $CT^s$ is the slim version of $CT$, $ID_{publish}$ is the ID of the $Publish_{CT}$ transaction previously submitted by the client. $PubID_{masternode}$ is a set of storage miners (i.e., their public keys) that are committed to store the data of $CT$, such that $|PubID_{masternode} |>r_s$, where $r_s$ is a protocol parameter added to ensure that there are enough copies of $CT$ in storage. 
\item $Solve_{CT}=<PubID_{miner},Sig(Sol),Score, \\ PubID_{CT},ID_{storage}>$, where $PubID_{miner}$ is the public key of the submitting miner, $Sig(Sol)$ is a digital signature of the submitted solution that acts as a commitment of the miner that his real solution will have the score $Score$. $PubID_{CT}$ is the identifier of the associated $Publish_{CT}$ transaction. $ID_{storage}$ is the public id of the storage miner from which data were retrieved.
\item $Validate_{CT}=<E_{dTMN}>$ is a transaction in which each member of the dTMN of $CT$ adds an encrypted list of past $Solve_{CT}$ transactions that were successfully validated.
\item $Seal_{CT}=<D_{dTMN}>$ contains the decryption specification for interpreting the $Validate_{CT}$ transaction. By knowing this information the winner of the competition is declared. 
\end{itemize}

\textit{Competition parameters}. Let $BC={B_1,B_2,…}$ be the ordered set of blocks in the blockchain. Denote $BC_{(i,j)}={B_i,…,B_j}$ to be a subset of $BC$ starting at block $i$ and ending in block $j$. A computational competition $CC_{CT}$ is associated with a single computational task $CT$ and has five phases: (0) store, (1) freeze, (2) compete, (3) validate, and (4) seal. Each phase has a defined number of blocks that mark its duration. We denote this number as $NB_x$, where $x$ is one of the competition phases. Let $I(Tr)$ be the block index of a transaction $Tr$ in the blockchain $BC$.

\subsubsection{Competition flow}

A computational competition acts as a microenvironment in our system and is illustrated in \textbf{Figure 2}. It lives as a set of related transactions on top of the blockchain. A competition over a task $CT$ starts when $Stored_{CT}$ is added to the blockchain. This is defined as a part of the protocol and every user or a miner that observes the addition of $Stored_{CT}$ to the block $B_j=I(Stored_{CT})$ can now compute the duration of each phase of the competition. \\

\begin{figure*}
  \centering
  \includegraphics[width=160mm,height=8cm]{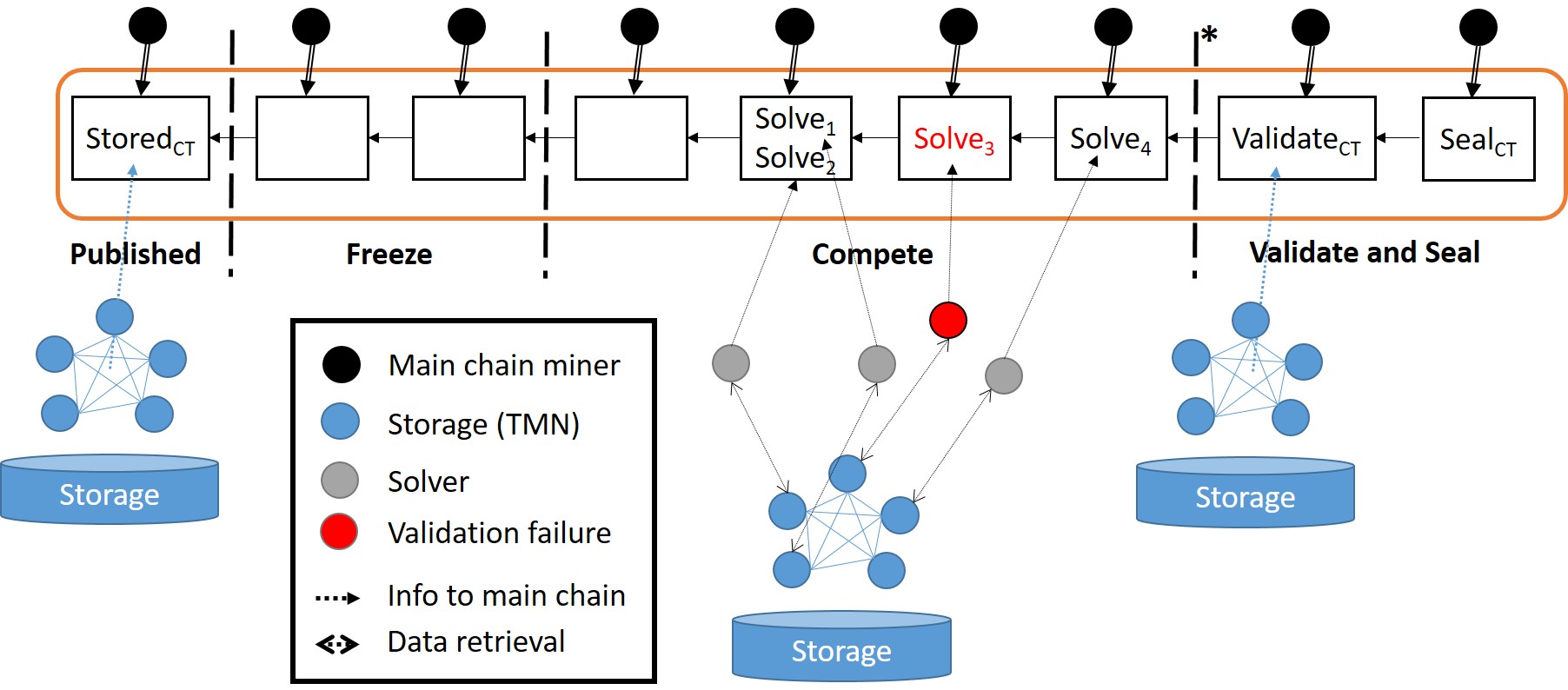}
  \caption{Illustration of the competition. Storage dTMN publishes the task to the network. After a freeze phase solving miners (gray) send out their solutions. In the validation phase the dTMN gets the solutions from the digital commitments given in the competition phase. In this case, the third proposed solution is marked as irrelevant (shown in red) in the validation transaction due to failure to fulfill the commitment. The asterisks marks that the validation phase may take a few blocks for the dTMN for reaching a consensus. Finally, a sealing transaction is added and profits are distributed.
}
  \label{Figure2}
\end{figure*}

Addition of $Stored_{CT}$ to the blockchain requires a pre-competition process that establishes replicates of data in storage and is fully described in \textbf{Section 2.4}. Briefly, this process starts when the client submits $D$ and its accompanying information to the storage chain in a transaction called $Publish_{CT}$. After a certain amount of time $D$ has a required set of copies in the storage chain, one copy per miner. This set of miners becomes the dTMN: a subnetwork of peers that are tasked with keeping $D$ and later validating proposed solutions. Throughout the competition, the activity of the dTMN is similar to a DASH maternode network. The peers continuously send out ping messages notifying the network that they are active. After validation has ended, which is through a consensus reached by the dTMN, they share a certain percentage of the fees that $CT$ entails. 

The next phase is freeze: from block $I(Publish_{CT})$ to block $I(Publish_{CT})+NB_{freeze}$ no $Solve_{CT}$ transaction is allowed into the blockchain. This step is one of our tools against an $O(1)$ attack. In this case, the attacker has no guarantee that her predefined solutions will be immediately profitable. Moreover, other miners will now have time to solve the published task. Next, the compete phase is the main place in which the blockchain stores proposed solutions for $CT$. For ensuring security of their solutions the solving miners do not send out their solution. Instead they present a digitally signed commitment. The reason for this commitment is to avoid a case in which the block signing miners will present the observed solutions as their own.

An important point in our protocol is that all PF fees are accumulated across sealed competitions, each in its designated pool, and are distributed every $B_{distr}$ blocks. We maintain three profit pools that are designed to reward the infrastructure components of the competitions: (1) the main chain (i.e., the public ledger), (2) the storage, and (3) the storage-main chain interaction. Note that the last pool is a derivative of pool (2): it is a defined percentage $P_{SMPool}$ of sums that the storage pool pays the PoW miners for accepting the storage infrastructure transactions into the ledger (e.g., bid, ask, or deal).

For every pool, funds are distributed in a way that is proportional to the amount of work  that was done by the miners, which can also be weighted by the transaction importance. For the main chain this is kept in expected value (i.e., on average over a stochastic process), whereas the other pools give out rewards in a deterministic fashion that is directly proportional to the amount of work done. For simplicity, we next discuss the algorithm of the main chain pool, which is based on enumerating the $Solve_{CT}$ transactions from the winning miners. The algorithms of the other pools are very similar but they are based on enumerating all relevant transactions in the ledger and not only those that are related to the winner of the competition.

Let $PF$ be the total sums in the main chain pool from the promised fees of the sealed competitions during the block range $B_i,…,B_j$ such that $j-i+1=B_{dist}$. We assign a weight $w_k$ for each block $B_k$ such that: $$\sum_{k=i}^j {w_k} =1$$. The weight $w_k$ is set to be proportional to the number of $Solve_{CT}$ transactions in $B_k$ that won their competition. The weight allocation algorithm is given below. For a block $B_k$ with a weight $w_k$ let $S_1,…,S_u$ be the set of $Store_{CT}$ transactions in $B_k$ that won their competition and let $M(B_k)$ be the miner that signed $B_k$. Given the output of the algorithm below, $w_k$ $PF$  is given to $M(B_k)$. \\ 

\end{multicols}

\RestyleAlgo{boxruled}
\LinesNumbered
\begin{algorithm*}[H]
\DontPrintSemicolon
\KwData{$BC_{(i,j)}$ am ordered subset of the blockchain }
\KwResult{$blockProfits$ with the weighted profit per block}
\Begin{
  \caption{MainChainWeightAllocation\label{alg}}
 $sum \leftarrow 0 $\\
 $w \leftarrow $ [$\emptyset$ for $k$ in $range(j-i+1)$] \# Python-like creation of an array of $\emptyset$s \\
\For{$k$ in $0:j-i$:} {
     wins=getWinningSolves($BC_(i,j)$ [$k$]) \# get $Solve_{CT}$'s that won their competition \\
    \For {win in wins:}{ 
 		$sum\leftarrow sum+getPF(win)$ \# Half the promised fee of the competition of win \\
 		$w$[$k$] $\leftarrow$ append($w$[$k$],$getWinnerID(win)$) \\
}
}
$W_t=getTotalCardinalWeight(w)$ \\
$blockProfits \leftarrow$ [$0$ for $k$ in $range(j-i+1)$]\\
\For {$k$ in $0:j-i$:}{ 
  $blockProfits[k] =  \frac{|w[k]|*sum}{W_t} $ \\
}
Return blockProfits
}
\end{algorithm*}
\hfill \newline

\begin{multicols}{2}

The next step of a competition is validation. Validation entirely depends on the dTMN, whose members wish to perform work, reach consensus, and seal competitions to gain more profits as discussed above. The storage miners here receive means to go back and validate the solution commitments submitted during the competition. The validate transaction holds for each dTMN member an encrypted list of $Slove_{CT}$ transaction ids that were successfully validated (for example, we can use hash functions). Upon publication of a validation transaction the dTMN publishes the last transaction that seals the competition: $Seal_{CT}$. Here, the dTMN members present the decryption algorithm for extracting their results in the $Validate_{CT}$ transaction. This defines a consensus solution and the competition winner(s) as the optimal solution that appeared first in the ledger wins. 

To emphasize the importance of maintaining pools that reward competition infrastructure we informally present a proposition below (elaborated in \textbf{Section 3}). Our goal is to illustrate the feasibility of an equilibrium in which it is the miners' goal to complete as many competitions as possible. 

\prop{\textbf{(Compete Nash Equilibrium feasibility)} Assume that the mean block reward per solve transactions is higher than the mean of regular non-competition transaction fees. The competition rules constitute an equilibrium in which the best strategy of hash block signers is to gather as many $Solve_{CT}$ transactions as possible, whereas computational task solving miners’ best strategy is to send their solutions. The optimal strategy of the storage miner is to answer retrieve requests of computational miners and send them the data. The optimal strategy of the block assembly miners is to enable $Solve_{CT}$ transactions to be added.}
\pf{Block signing PoW miners know that by including more solve transactions that have they increase their expected relative weight in the output of MainChainWeightAllocation. Thus, when they assemble blocks the best strategy for increasing the expected profit is to include the solutions with the best declared scores. The exact number depends on three parameters: (1) how many competitions are alive, (2) the difference between the mean reward of standard transactions and solve transactions, and (3) the probability that the submitted commitment will fail.

Storage miners are paid proportionally to their work. They would like to have as many computational miners as possible whose data copies were retrieved from them. On the other hand, the computational miners, which aim to win competitions by submitting their results, know that the submission associated fee is a PF. Thus, they only pay it in case they are selected to be winners and therefore they have incentive to submit solutions \qed}

\subsubsection{How to keep the solution safe}
A natural question regarding the competition protocol above is how can we make sure that upon sending a solution an attacker cannot take it and claim it as is own? We propose two options to solve this issue, see \textbf{Figure 3}. 

The first solution is based on the idea of cryptographic commitment. This is the digital analog of taking information, sealing it in an envelope, and putting that envelope out on the table where everyone can see it. This means that the solver commits to the score written on the envelope, but the solution remains a secret from everyone else. Later, in the sealing phase, the winner can open the envelope and reveal the value that he had committed to earlier. A simple implementation of commitment can be done by utilizing a hash function: the miner sends the hash of the solution and a nonce. After the competition is over the winner can prove his commitment by sending the real solution together with the nonce to the blockchain in order to finalize and seal the competition.

\begin{figure*}
  \centering
  \includegraphics[width=150mm,height=7cm]{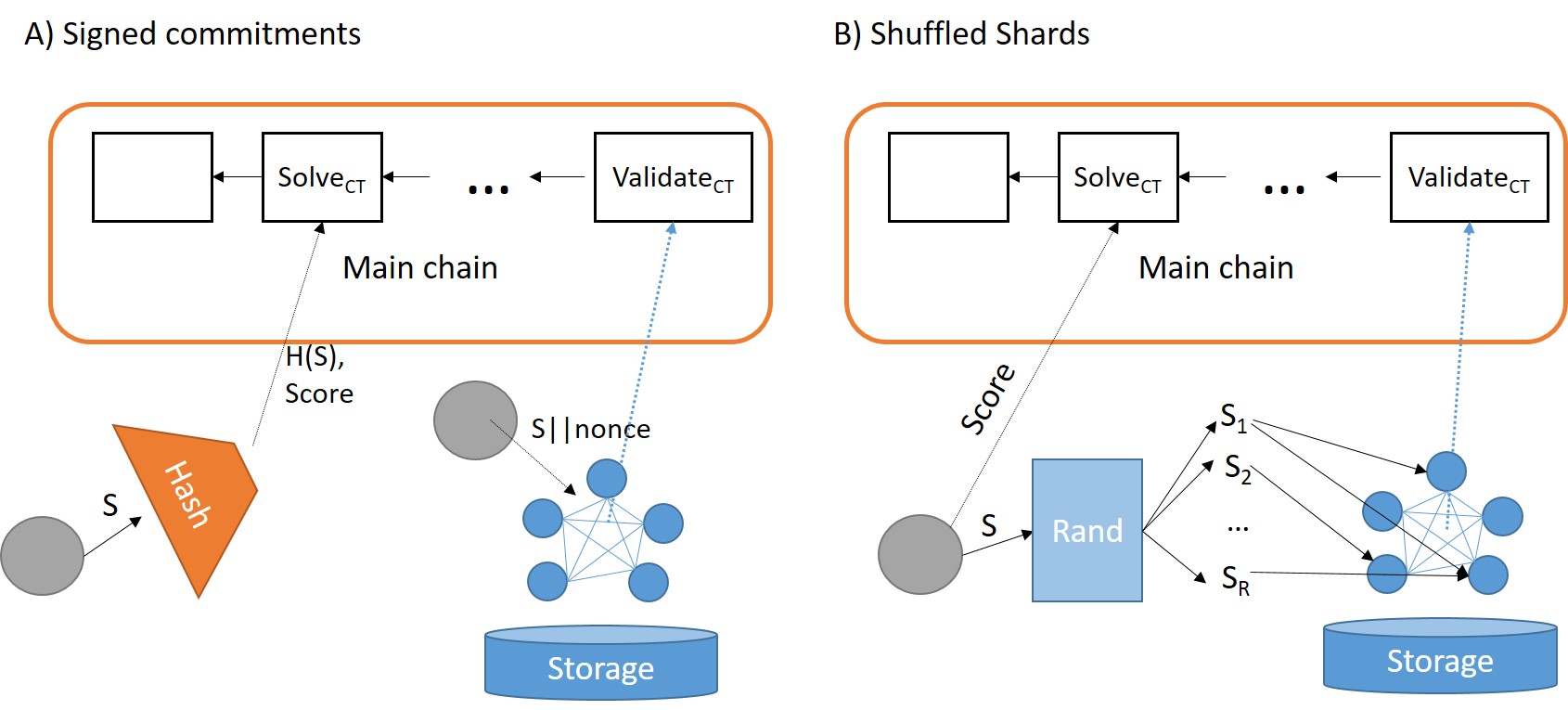}
  \caption{Two digital mechanisms that can be used for signing solutions. A) Standard digital signatures require the miner additional activity after the compete phase as only then the dTMN require the solution. B) Shard based method does not require a second interaction with the solving miner but it puts all trust on the dTMN.}
  \label{Figure3}
\end{figure*}

Another solution is based on randomly partitioning the solution into shards. Here, the miner submits a few shards for each member of the dTMN. Each shard has a randomly assigned number, taken at random from a large space that the miner does not share with anyone. Only when all shard numbers are visible can a dTMN member order the shards and assemble the correct solution. This solution removes the need of the solving miner to stay alive after a competition, but substantially increases the dependence in the dTMN.

\exa{\textbf{(Random shards)}Assume that the miner partitions the textual representation of the solution into $12$ parts $S_1,S_2,…,S_{12}$. Then, each shard receives a random number between $1$ and $100,000$, for example:
\begin{eqnarray*} &100,500,985,12354,50000,60000,	60001,69001, \\
& 80000,90000,90020,999000\end{eqnarray*}
If two shards are randomly selected for each member of the dTMN, and assume that there are seven nodes in the dTMN then each node holds $2/7$ of the solution and with high probability does not even know if the shards they have are close or not in the correct ordering. When it is time for validation, the miners exchange shards. Only if a piece of the solution is missing the nodes contact the computational miner. 
}

\subsubsection{Block sizes and transactions}

Given that a computational competition requires several transactions, going from the pre setup phase by the storage chain (discussed next) to the sealing phase through miner-submitted solutions, we need to make sure that current block sizes can enable the new protocol. \textbf{Figure 4} shows the block sizes of Bitcoin and Ethereum over time. Bitcoin’s block size has been stable at near $1$ mb per block rate of $10$ minutes. Ethereum’s blocks are largely $>20$ kb where the average block rate is below 20 seconds. \textbf{Figure 5} shows statistics for the number of transactions for the two networks. Given that: (1) Decred aims to keep its blocks at around $1$ mb, and (2) our transactions will be light thanks to keeping most information in the storage chain we conclude that the burden of maintaining competition is within the capabilities of current block sizes.

\begin{figure*}
  \centering
  \includegraphics[width=130mm,height=80mm]{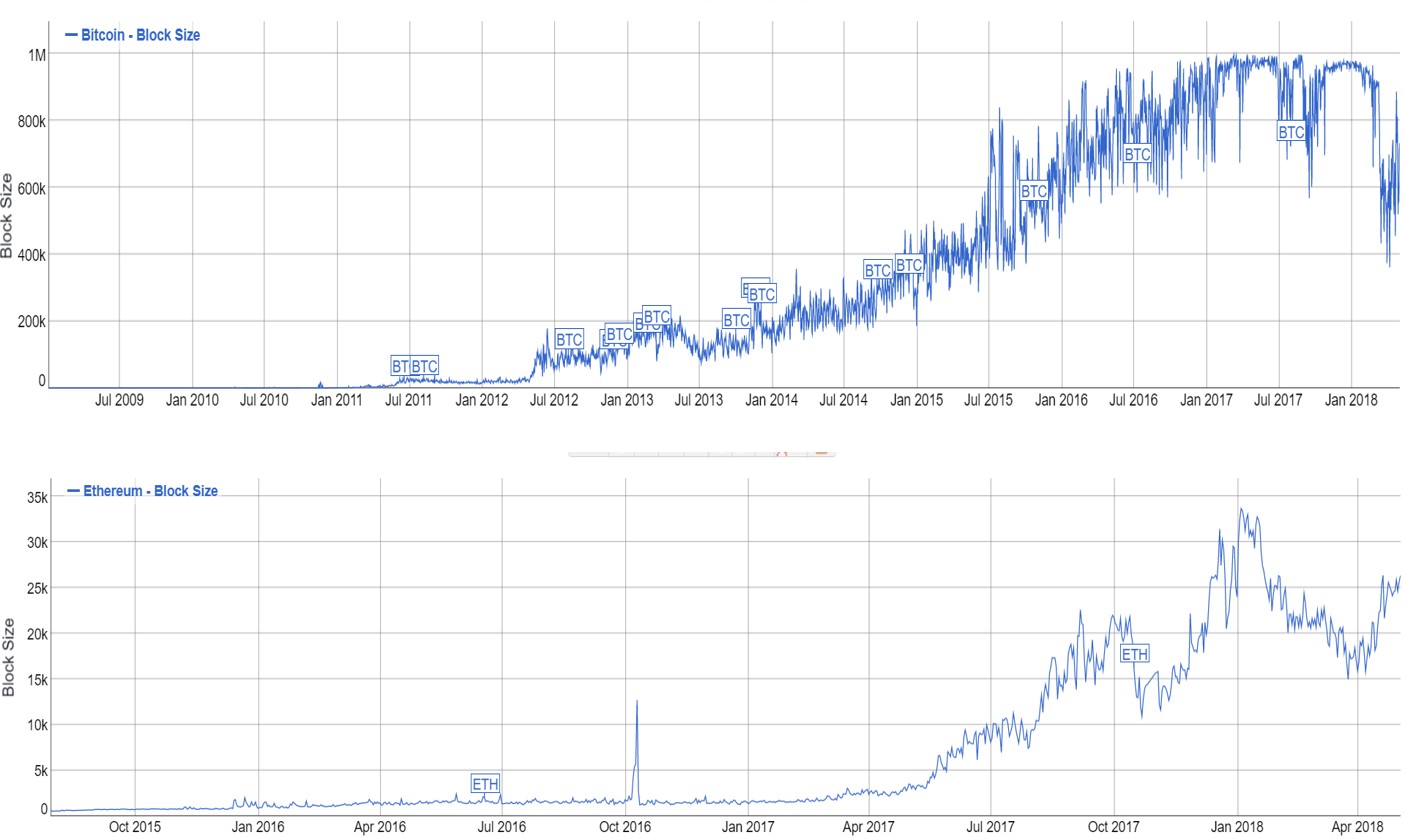}
  \caption{Block sizes of Bitcoin and Ethereum. Source: https://bitinfocharts.com/}
  \label{Figure4}
\end{figure*}

\begin{figure*}
  \centering
  \includegraphics[width=140mm,height=140mm]{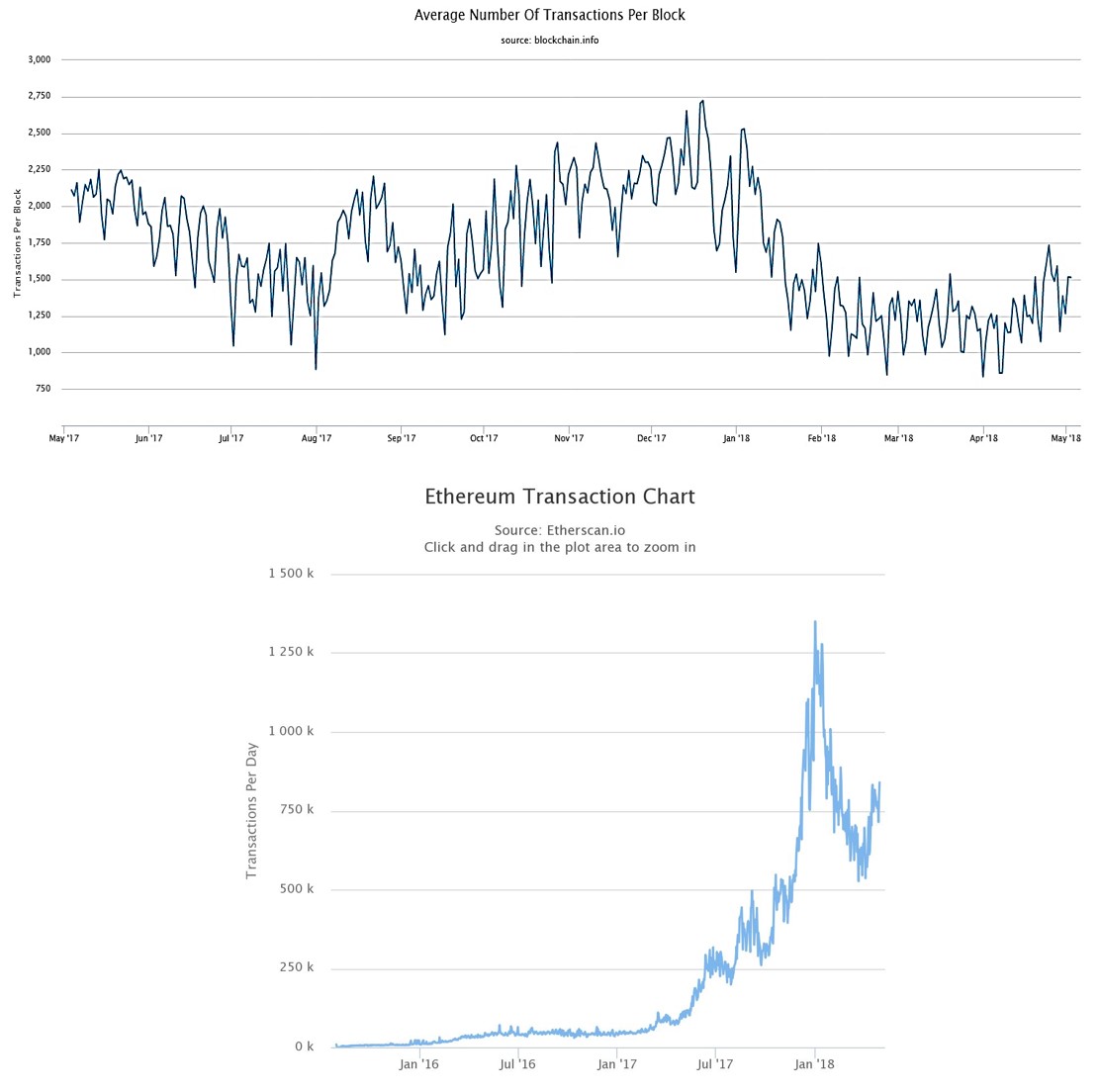}
  \caption{Number of transactions in Bitcoin and Ethereum. Bitcoin: number of transactions per block. Source: https://blockchain.info. Ethereum: number of transaction per day. Source: https://etherscan.io/chart/tx. }
  \label{Figure5}
\end{figure*}

\subsection{The storage chain}

Our storage component can be easily built upon existing storage blockchains. Roughly, these systems address the following issues:
\begin{enumerate}
\item Closing a deal: create a reliable platform that enables a clients market and a miners market
\item Payment for storing the data 
\item Proof-of-Replication: how miners prove they hold a valid replica of the data
\item Payment for retrieving the data
\item Data encryption (may be if left to the users)
\item General blockchain maintenance
\end{enumerate}

Our requirements are only a subset of what these systems address. It is crucial however, that off-chain micro-payment will be available: when two parties transfer data we use a protocol that ensures the transfer but keeps most of the information off chain (proofs and certifications are kept for tracking). As exemplary systems that provide all functions above we consider Filecoin \cite{ProtocolLabs2017} and Storj  \cite{Wilkinson2016}. For a thorough explanation on each system, see the \textbf{Appendix}. 

\subsubsection{Our adapted storage protocol}

Clearly, one can use any storage to keep important task data. In fact, we can even consider using centralized cloud services such as Amazon or Google Cloud. However, the entire protocol would become centralized and tightly dependent on an external provider, which contradicts our goal to create a viable self-contained system. Technical side-effects of using external providers include: (1) the network is now sensitive to the provider's API modification and maintenance, (2) using an external coin requires additional exchanges, and (3) the system has no control or effect on the storage cost.  Another issue is inefficiency in that: (1) data will be copied twice for validations, and (2) it will not exploit the fact that miners that hold the data can easily validate solutions and there is practically no added heavy cost to the hardware they need to keep. The dTMNs (task-specific masternode networks) take upon themselves the maintenance of the task data and the validation of the solutions once the competition phase is over.

Our default suggestion is to take Filecoin as base code, change it for our needs and make the storage \''smarter\'' by supporting validations. 

\begin{flushleft}\textbf{Storage constraints and collaterals}\end{flushleft}
In this section we adopt standard storage-based notation that uses time span instead of counting blocks to allow more granularity. Below we define parameters and their constraints.
\begin{itemize}
	\item Let $T_x$ be the block number at time $T$ auditing the transaction $x$.
	\item Let $timeout_x$ be the number of blocks for timeout transaction $x$.
	\item $bid$ request becomes invalid after $timeout_{publish}$ blocks starting from $T_{publish}$ 
	\item A miner that publishes an ask request in $T_{ask}$ must lock the needed resources for $timeout_{publish}+T_{bid}-T_{ask}$
	\item The dTMN of CT promises to store the dataset for at least $timeout_{freeze}$ starting from $T_{stored}$
\item After $T_{solved}$ storage miners must keep the data for a $timeout_{retrive}$
\end{itemize}

\begin{flushleft}\textbf{dTMN problem gathering}\end{flushleft}

The client starts by submitting a bid request (in-chain) in a transaction called $Publish_{CT}$. Then, storage miners submit ask requests. Only after at least $r_s$ miners publish an ask request, the client and the miners sign joint deal transaction, which we previously denoted as $Stored_{CT}$ (see \textbf{Figure 6}). Using a multi-signature process, this group becomes the dTMN that provides services. All subsequent transactions, including those providing PoSt, are made as a group via one multi-signature transaction. 

\begin{figure*}
  \centering
  \includegraphics[width=90mm,height=50mm]{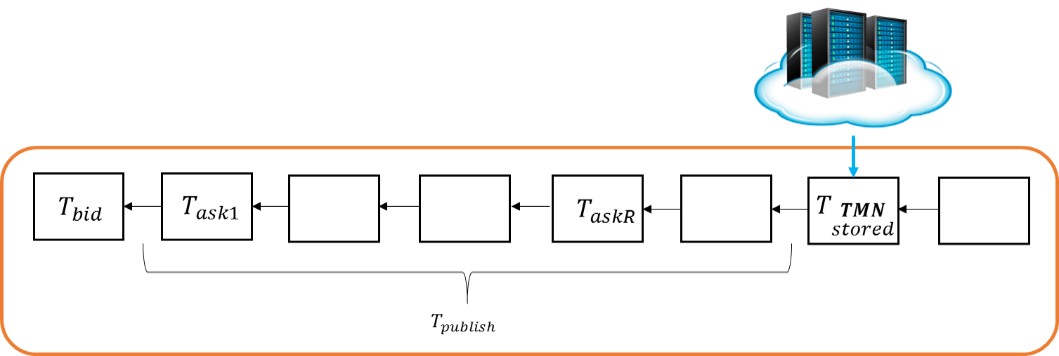}
  \caption{A simple illustration of the process that leads to a Stored transaction. }
  \label{Figure6}
\end{figure*}

In order to minimize the number of in-chain transactions, storage services are audited using micro-payment method. Micro-payment channel is used by storage miners and computational miners, as well as with the client on solution retrieving. Proofs of micropayments (e.g., claim micropayment transaction) are done once by the dTMN after the computational competition process is completed. The same holds also for PoSt proofs. Having said that, a single node of the dTMN may also submit a transaction when the dTMN parties are not honest. dTMN may become smaller due to parties failure, therefore the network keeps following the dTMN size. If the dTMN gets too small (e.g., less than  $\frac{r_s}{2}+1$) then the network considers this dTMN as failed and the computation process is stopped.

The dTMN, as an efficient storage network unit will send and receive data in chunks. In order to keep high-availability, each node in the dTMN holds the whole copy of the data. This is achieved by chunk exchange. Chunk exchange is highly beneficial here because it is much faster than sending the whole dataset to each miner.

\section{Protocol properties}

\subsection{Forging via 51\% attacks}

Carrying a 51\% attack in our system, by definition, requires achieving 51\% of both the PoS and the PoW components. This is a useful property of Decred: hash solving miners propose new blocks, whereas PoS miners approve past blocks. An adversary that reaches only 51\% in the hash component still needs to get the proposed blocks approved. An adversary that holds 51\% of the stakes cannot suggest new blocks. Thus, when there are many miners of both types, it becomes very difficult and resource consuming to try such attacks. This is a valuable property is not violated in our system thanks to the accumulation of past useful work. 

\subsection{$O(1)$ attacks}

The competition's freeze phase makes $O(1)$ attacks that involve low difficulty problems unlikely because the adversary has no guarantee that he will be able to win his own offer back. As a result, $O(1)$ attacks with notable outcomes require acquiring many coins, submitting problems with a very high difficulty, which will also result in locking them for long time periods. This is, in essence, equivalent to performing PoS because the adversary first needs to acquire coins in order to submit a new task (a client that has no coins cannot offer $fee_{solve}$), and then these coins are locked until the competition rewards are distributed after $>B_{dist}$ blocks. On the other hand, our \textit{PoS higher profit invarnace} guarantees that the $O(1)$ adversary is better off with performing PoS, which does not even require computational resources.

Under the constraints above we are still able to set the protocol parameters such that the system the honest miner that won a competition gets more than $fee_{solve}$, submitted by the client. Why is this even possible? Keep in mind that we have a gap: using vstakes to purchase tickets is not equivalent to using money to buy tickets. The former is not entitled to the extra rewards that the pure PoS miner expects. This gap allows earning $>100$\% compared to the client-submitted fees, as we shall show next. 

\thm{For a ticket price $y$ there exists a set of parameter assignment to $r,p_{pools},p_{vstake}$ such that the expected profit of a pure PoS mining is higher than that of an $O(1)$ adversary.}
\pf{

Assume that the adversary submitted a problem with an offer to pay $y$. By definition, $p_{pools} y$ is paid to the pools that reward the storage and main chain infrastructure. He then gets $y(1-p_{pools})$ back by solving his own problem and $p_{vstake} y(1-p_{pools})$ in virtual stakes. These are minted new coins and there is no added reward. The total reward of the adversary is:

\begin{eqnarray*}
&y(1-p_{pool} )+p_{vstake} (1-p_{pools} )=\\&(p_{vstake}+1)(1-p_{pools} )y
\end{eqnarray*}

Performing PoS in the same time by buying a ticket in price y would have resulted in an expected reward of:

\begin{flalign*}
& Pr(ticket \ is \ selected)ry \ + \\
& yPr(ticket \ is \ not \ selected) \ = \\
& 0.95ry+0.05y=y(0.95r+0.05)
\end{flalign*}

Setting $r=1.1$ we get an expected pay back for PoS of $1.095y$.  We can now easily set the other parameters such that:
$$1<(p_{vstake}+1)(1-p_{pools} )<1.095 $$
For example, $p_{vstake}=0.25,p_{pools}=0.15$ yields $1.0625$. 
Alternatively, if we want to mint less coins per virtual stake then: $p_{vstake}=0.15,p_{pools}=0.1$ yields $1.035.$ \qed
}\\

We proved the theorem above by example. However, note that the analysis above defines an infinite number ways to select parameters. To see this, assume we wish to keep a given percentage $\epsilon_{pos}$ difference between the expected PoS reward and the reward of a computational miner. Assume we wish to maximize the computational miner’s profit. This amounts to the following constrained problem:
\begin{eqnarray*}
& argmax_{p_{vstake},p_{pools}} (p_{vstake}+1)(1-p_{pools} ) \\
\end{eqnarray*}
s.t.
\begin{eqnarray*}
& (1) \ \ \ \  (0.95r+0.05)-\epsilon_{pos} = \\
& (p_{vstake}+1)(1-p_{pools} ) \\
& (2) \ \ \ \  0\le p_{vstake},p_{pools} \le 1
\end{eqnarray*}

\textbf{Figure 7} below illustrates how for $\epsilon_{pos}=0.01$ we get an infinite set of ways to adjust the parameters for different expected PoS rewards. \\ 

\begin{figure*}
  \centering
  \includegraphics[width=80mm,height=70mm]{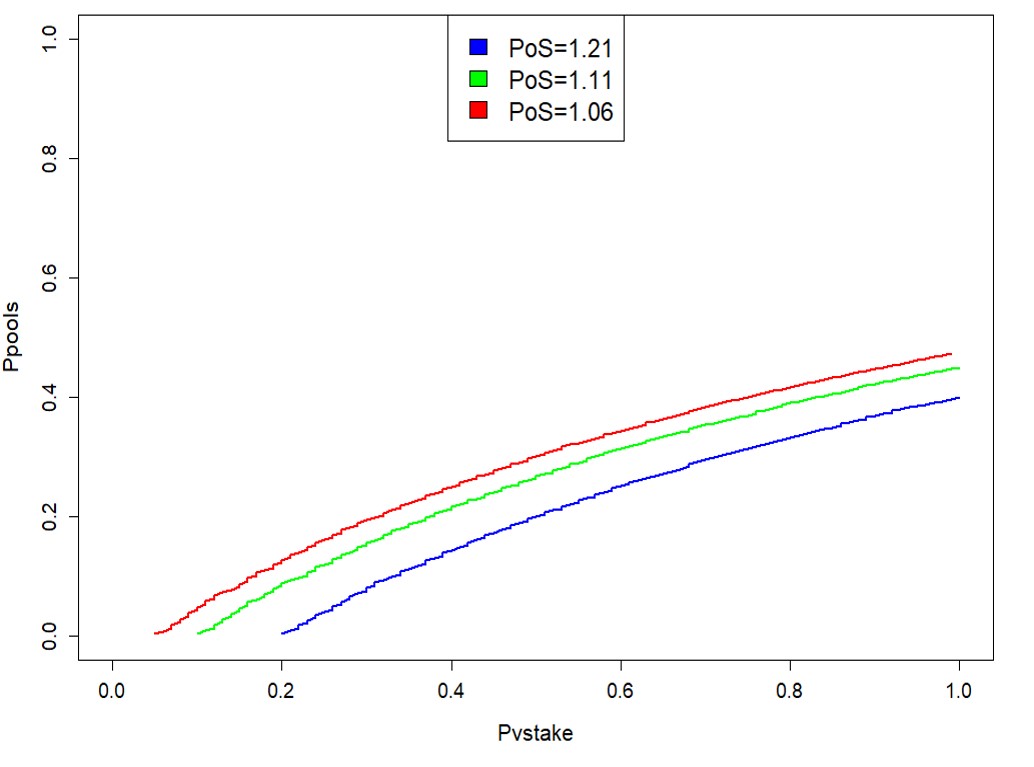}
  \caption{Computational miner reward parameters under the PoS constraints. For different PoS expected reward factors and a desired margin of $\epsilon_{pos}=0.01$ each line shows the set of $p_{vstake}$ and $p_{pools}$ combination that satisfies the constraint in which PoS is more profitable than $O(1)$ attack.}
  \label{Figure7}
\end{figure*}

\subsection{Suboptimal solutions}

The computational miner can submit many solutions but the system discourages sending out too many suboptimal solutions, under the assumption that there are honest miners working on the same problem. An honest miner that did substantial work is likely to offer a small transaction fee. This is crucial if the PoW miners are bombarded with solutions. In this case, the honest computational miner looks at the ledger and knows that his solution is better (because solution scores are not encrypted). He is then likely to offer a transaction fee that is similar to that of an average non-competition transaction fee. In this case the SSA (suboptimal solutions adversary) that tries to overwhelm the system with solutions will have to pay substantial funds. This concept is similar to early suggestions for fighting spam emails: the spammer needs to pay a lot to succeed, whereas the honest users pay negligible sum per transaction \cite{Dwork1992}. Moreover, note that the PoW miner will also accept solutions with better scores into the block as these increase the chance to get sums from the pools. Thus, the SSA knows that submitting random solutions is a strategy that is likely to end in loss. 

\subsection{The dTMN and fault tolerance}

Byzantine fault tolerance (BFT) is the dependability of the system in failure of components (and the identification thereof) \cite{Kirrmann2005}. We discussed above several mechanisms to prevent colluding attempts (e.g., make
storage miners be more like DASH masternodes). However, for BFT, note that our system always depends on both reaching a consensus and getting it approved. For the competition validation phase we give vote management to the block miners by letting storage miners submit commitment of their votes. Thus, there is no need for dTMN members to coordinate their response and in case some miss their chance due to failure, there are still others that can reach consensus. The dTMN members are expected to send their results separately but consistently until the validate and seal transactions are published. This is similar for requesting masternode members or PoS right owners from other protocols to be present in a given time – otherwise they lose their rights.

\subsection{Block withholding attacks}

Block withholding, also called selfish mining, is an attack in which a large mining pool withholds mined blocks in hope to mine another one and let the network pursue an orphan block. By doing so, the attacker can present its longer chain later and make the network accept it as the new consensus \cite{Eyal2014}. The existence of this vulnerability and proven strategies that incentivize miners to join such pools, shows that the Bitcoin protocol in fact may promote decentralization by joining to pools, that in turn can attack the system even with as low as 25\% power (after the improvements suggested by \cite{Eyal2014}). Several analyses had deepen the understanding of such attacks, but were mainly confined to Bitcoin itself or to other very similar protocols \cite{Bag2017,Sapirshtein2017}. 

Nevertheless, this attack still requires large computing power, and is currently not suitable for hybrid PoS systems, especially ones similar to Decred in which the PoS approves blocks. Here, withholding a block and letting the network focus on a shorter chain is in fact not beneficial once the PoS mechanism approved the shorter chain. In these cases, the shorter chain is accepted as the consensus because it is the only one whose block before the current head was approved. Even though our system aims to deviate PoW power to the computational tasks and thus reduce the hash difficulty, it does not come at the expense of this property.

\subsection{Can miners collude?}

Given that the computational task miners all compete for rewards from the client, there is a certain risk of colluding. For example, miners may agree to send out random solutions without doing any work. Note however that we only need a \textbf{single honest miner} per computational task to render the entire colluding effort barren. This property weakens the conspiring adversaries substantially as they do not necessarily know that all miners had agreed, which is very difficult if the network is large. Moreover, even under such agreement the entire system is still under the prisoner's dilemma: even rational colluding miners are not likely to cooperate especially if they know that all they need to do is a relatively small amount of useful work to win the competition and claim the rewards. Thus, betraying the partners results in a greater reward. As all miners are aware of this, the rationale behavior will be to start competing, especially once a non-random solution score is visible on the blockchain.

Finally, we can consider active mechanisms that can be added to prevent such attacks (if required in practice). First, in a similar fashion to the way public machine learning competitions are held, the organizing institution typically performs a simple initial analysis of the task and submits the result to the network. Selecting tasks at random and submitting the solution at a random time point dismantles the attack and provides a reasonable solution to the costumer. Second, having a public reputation board will be useful in any case. However, this is a more involved solution that will require a thorough examination of the incentives.

\section{The agents and the market}

The goal of this section is to review the protocol from the viewpoints of the agents of the system and describe how the market will play a role in adjusting the incentives and the competitions. This discussion is important because useful work systems, except those designed solely for contributing hardware to specific goals (e.g., SETI@home) \cite{Korpela2001}, depend on the clients. 

\subsection{Computational miner's view}

\subsubsection{Choosing a problem for solving}
A computational miner can choose any stored computational task that has a matching $stored_{CT}$ transaction (but not $validate_{CT}$ nor $timeout_{retrive}$ expired). Retrieving a problem credits the storage miner in a way that is proportional to the associated fee. This fee can be later used as a part of the storage pool from which storage miners receive additional rewards as explained above. Technically, this is promised by using micro-payment protocol between the storage miner and the computational miner whose final output is shown in a deal transaction in the main ledger.

\subsubsection{Publishing solutions}
Publishing a solution is allowed only after the freeze phase. Using the freeze phase serves two goals: (1) it gives time for the miners to evaluate whether or not they have the required resources to join the competition and (2) it prevents $O(1)$ attacks that overwhelm the system with easy tasks. The solution is stored by the dTMN and a signature of the solution is published in-chain by $Solve_{CT}$. It is important to note here that even though the storage has indirect incentive to reach to validation phase, the storage miners are directly incentivized because sending out data credits them when pool sums are distributed. In \textbf{Section 2.3} we suggested two ways of storing a solution such that the miner’s work is not compromised: (1) Signed commitments, (2) Shuffled Shards.

\subsubsection{Getting more than solving fees}

A natural point against useful work within a decentralized blockchain may be the following. If the client pays $a$ and the system adds $b$ then the total cost of a task for the system is $a+b$, whereas the miner is entitled only for $c<a$.  Such analysis may be suitable for past useful work suggestions. In our system this point is avoided thanks to two main considerations. First, as we showed above the computational miner receives $c>a$ without opening the system to $O(1)$ attacks. 

Second, we must consider the balance between the computational tasks market and the blockchain market (including a minting process, if used) to correctly analyze the effective cost of the task. Generally, the client first pays directly to miners that solve tasks and for the blockchain infrastructure. Note that these fees are indirectly under the market’s regulation (e.g., by comparing to cloud service costs): these prices must be appealing for the miners as compared to solving hash puzzles. Making miners solve tasks diverges the system towards solving useful work instead of hash puzzles. Thus, the same energy that was otherwise used by the system to solve harder hash puzzles is used for tasks and there are no expenses incurred. 

So, where does this extra wealth is coming from? The answer is the blockchain itself, if its own market has a value. That is, the set of products that can serve financial systems and enable trading.

How does PoAW serve the blockchain? Our answer here is simple: it is accumulated and used for the PoS system, which in our case is also useful for improving governance and avoiding N@S issues. Thus, as implied in \textbf{Section 3.1} above, these accumulated stakes are used to strengthen the system such that if only a small group of miners chooses to solve hash puzzles, they cannot carry 51\% attacks. 

\subsection{Storage miners}

\subsubsection{Storage verifies the solution}
After freeze the phase is over, the solution can be revealed.  At this point each node of the dTMN verifies all submitted solutions. Then, miners start the voting process in order to reach consensus. The dTMN publishes a multi-signed validate transaction $Validated_{CT}$  containing all voting results, encrypted. One can think on the following voting strategy: vote as the current majority. That way, he doesn’t have to validate the results. Therefore, each vote is encrypted, and only after publishing the encrypted voting results, the keys are added to the $Seal_{CT}$ transaction. We have therefore reached a dual result. First, now dTMN members have no incentive to cast erroneous or random votes. Second, the protocol can declare a winner that is the first validated best solution.

\subsubsection{Network payments and incentives}
Payment to the storage will be done only after the solution was kept for some time factor (letting the client pull the solution). Only storage miners that validate the answer correctly (vote as the majority) on all the solutions will be paid. Per storage service, the storage miners are paid from the storage infrastructure pool. Their fees are extracted automatically only for services that were proved. As a result, each storage miner of the dTMN will be payed per given service. The payment is proportional to the amount of his work. For example, if a storage miner uploaded $X$ kb and the total uploaded kb of the storage in the analyzed time window is $Y$, then he will be paid by $X/Y$. Note that for receiving and storing the data from the client the dTMN members are paid directly not from the pool. Finally, note that there is no need to further pay for validation as without depositing the validation results, the competition cannot end and the payments to the pools will not occur. Our method encourages storage miners to do as much work as possible per computational task as a part of their optimal strategy (see Proposition 2.3).

\subsection{The clients and the market}

\subsubsection{Network maintenance costs}
Upon submitting a computational task, the user specifies a set of fees and promised fees:
\begin{itemize}
	\item $fee_{tr}$: same as any other transaction (payment to the main chain for taking the transaction)
	\item $fee_{sub}$: a bid in the Storage Market, used for paying for storage
	\item Retrieval: retrieving data in the Retrieval Market, used for paying the storage for sending data
	\item Validation: used for validation payment
	\item Maintaining: used for paying the main chain for holding all other transactions
	\item $fee_{solve}$: the payment to the winning solver
\end{itemize}
Note that the promised fees above can be predefined percentages of $fee_{solve}$. Moreover, the protocol can define a set of constraints, e.g., that the total sum of payments to the main chain pool is at least 15\%. Optimal assignment of these parameters is still under research.

\subsubsection{Client retrieves the solution}
After voting for the best solution, the client can retrieve the data from storage (off-chain) using micro-payment protocol. The dTMN will be paid only after $timeout_{retrive}$ for letting the client time to retrieve. 

\subsubsection{Useful work difficulty}
Previous useful work suggestions aimed for replacing hash puzzles with a completely new mechanism. This created a problem with the need to correctly estimate the difficulty of each instance from each considered problem. For most problems, this task is currently not feasible as complexity analysis usually involves obtaining bounds for the worst-case or average-case running time. Our approach is radically different: we let the client decide how much each task is worth. This means that the clients have to evaluate the cost by comparing to other alternatives in the market and then suggest what they consider a fair proposal. Of course, this will need to be balanced between the goal of saving costs (compared to other alternatives) and attracting miners.

\subsubsection{Storage cost balance}
The market determines the balance solving fees and storage fees, which are proposed separately (do not confuse the initial storage fee from the downstream funds that go for maintaining the storage pool). This balance cannot be evaluated in advance theoretically as this is subject to fluctuating costs of hardware and energy consumption. We therefore, as in the previous section, let the market reach the optimal balance, which is also expected to be updated dynamically.

\section{Computational competitions}

In this section we discuss the merit of competitions and the fields that may benefit from them. Note that in this paper we discuss as an example having a competition for a hard optimization problem. In general, our methods can be adapted to additional competitions that are of better fit to certain fields such as machine learning (work in progress). For simplicity, we keep the discussion about optimization problems only.

\subsection{The system's efficiency}

Our protocol, like any decentralized blockchain protocol, is bound to have redundancy. First, the blockchain itself has many copies. Second, each member of the dTMN has a copy of the data of a task. Finally, the computational miners compete for gains from the same problems. The first two redundancies result from the need to reach consensus between agents in the system, which is crucial for decentralization. We therefore accept these two as needed constraints. The third redundancy,however, is a much more delicate issue and it is the main focus of this section. 

From the computation perspective, our suggested competition-based system substantially differs from the classic view of utilizing distributed systems for computational tasks, which are predominantly centralized systems. Moreover, many distributed computing protocols where designed for managing large clusters of many limited power nodes. In contrary, in our network each miner may be a powerful agent with substantial computing resources. This can be seen as having a network of clusters such that each cluster has a known manager \cite{Abed2006}. 

\subsubsection{The irrelevant WC}

To make our discussion more formal we compare our system to an optimal centralized system with no reliability or connection issues. Let $n_m$ be the number of miners in the system and $n_t$ be the number of tasks. We make the following simplifying assumptions: (1) $n_m$ and $n_t$ are the same in both systems (2) all miners have the same computing power, (3) all miners have the same same energy consumption per computing unit per time, (4) in our system miners select tasks at random, uniformly over all tasks, (5) all tasks require the same resources exactly, they take one hour to run in a single machine (to make time discrete and standardized), and (6) all tasks are all worth the same value $v$. Let $e$ be the ratio between the income per hour in the centralized system and the expected income in our system. 

Clearly, the worst-case scenario for our system is when there are only a few tasks and many miners solve the same problems using the same deterministic algorithm. In this case, the expected number of solved tasks in our system is $\frac{n_t}{n_m}$. Assume $n_t < n_m$, then:
$$e = \frac{v n_t }{v n_t / n_m} = n_m$$
which means that the system is inefficient in a way that is linear in the number of miners.

The analysis above is irrelevant for us as our goal is not to use our blockchain system to provide a general purpose work stations for users, nor do we seek to replace current centralized cloud systems. Alternatively, we seek to solve computing and modeling bottlenecks that are widely common in large optimization problems. These problems are typically solved using algorithms that either have stochastic component to them or there is a large set of hyperparameters that guide the search and there is only partial prior knowledge on how to set them. Examples for sources for internal randomness include: selection of an initial guess (e.g., SGD), random variable ordering (e.g., within a branch-and-cut algorithm), Markov Chain Monte Carlo (MCMC) simulations, and simulation within genetic algorithms. Moreover, in many situations such algorithms can be distributed into small tasks. This type of problems are the basis for machine learning and pattern recognition and we shall discuss specific examples below. 

\subsubsection{Closing the gap}

The efficiency gap illustrated above quickly vanishes when the user's perspective is taken into consideration. Most importantly, for almost all optimization problems for AI  one of the following is true. First, users need to rerun their tasks several times in order to examine how the internal randomness of the algorithm affects the solution. For example, in SGD, we may accidentally choose a bad combination of an initial solution and optimization parameters that will lead to a useless result. Second, the problem can be distributed into many small tasks that can run in parallel (this step may run sequentially, but each iteration involves parallelization). For example, in integer programming there are many algorithms that break the input problem into many subproblems (e.g., by taking a subspace of the variables).

The two points above directly affect $e$ in a way that creates a trade-off. First, distributing a large task into many small ones means that both the difficulty of each task can be decreased and $n_t$ can be much larger than $n_m$. In these cases, the overall income of the system depends on how many tasks can a miner take on. In any case, distribution can bound the number of miners per competition.

Second, the true cost for users is a function of the total number of times they need to run the algorithm in practice. Moreover, even with different attempts there is no guarantee that the solution converged to the best one (or even a useful one) and no guarantee that the search space was thoroughly examined. In contrast, using our competition-based paradigm there is a single run of the competition and the powerful miners compete by testing different options. Thus, the costumer can propose a much lower cost as compared to the effective cost (e.g., five times the price per task, assuming users will execute 20 runs in a centralized system) and miners will still see it as a very high profit opportunity. Moreover, the competition can even achieve a better final result compared to the centralized system in which the user have limited resources that reduce the number of attempts.

\subsection{Machine learning competitions}

The lack of one master algorithm for all machine learning or optimization problems is a known result, also known as the \textit{no free lunch theorem} \cite{Wolpert1997,Wolpert1996}. Empirical difficulties over time had led the community to establish competitions and challenges in order to achieve progress. These had proven to be extremely valuable for advancing machine learning. One of the most famous examples for this concept is Kaggle, which is a platform in which data scientists can compete for challenges posted by organizations (e.g., \cite{Kaggle2017}). Another, more academic example is the DREAM challenges in computational biology define competitions for specific applications. Then, by investigating and integrating the submitted solutions progress can be made, for example, by adjusting algorithms and their parameters \cite{Prill2011}.  

There are many examples for successful competitions in Machine learning and we cannot go over them all here. These competitions are especially useful for advancing applications that had proved difficult over time. We focus here on two examples that historically led to substantial progress. First, the Netflix challenge that led to a marked progress in Collaborative Filtering as tool for creating recommendation systems \cite{Wu2006}.

Second, and more famous these days, is Deepl Learning. This is a renaissance of the neural network machine learning branch that started roughly in 2012 thanks to the ImageNet competition \cite{Krizhevsky2012,JiaDeng2009}. In this challenge, competitors were asked to learn image classifiers for $>5000$ possible labels using $>3$ million images. A multi-layered \''deep\'' neural network achieved top performance by a substantial margin.  

Over the last years this field produced machine learning models that markedly outperform previous alternatives. The field has attracted an influx of interest from both academy and commercial companies. Unlike other statistical learning fields, the most important achievements are based on a purely empirical heuristic research. These methods are used in many fields today ranging from image processing and speech recognition to applications in quantum physics. The nature of the field also precludes using a limited number of models for all applications. This is exacerbated due to the reliance on numerical experimentation to select the best architecture for an application.

Our protocol can be used to establish a competition for each application in the blockchain as long as the underlying problem can be formulated as an optimization problem. Based on the empirical observation that competitions are a useful way to enhance machine learning, our system is expected to improve upon state of the art in many tasks. 

\subsection{Practical applications}

As a proof of concept we are developing pipelines for computational tasks to solve two important practical problems: deep learning and mixed integer linear programming. Simpler versions of these problems were presented in \textbf{Examples 2.3. and 2.3}. Our goal is not to provide a thorough review of these fields, nor do we aim to present novel methodology. In this section we present basic algorithms and illustrate how they can be used to solve problems using our system. Note however, that unlike our proof of concept below, our system is much more general as it does not impose on miners which algorithm to run.

\subsubsection{Deep Learning}

Training neural networks for a specific application requires several steps. First, we need to define the network architecture. This includes the number of layers and their types, which determine the internally extracted features. Second, hyper-parameters are set. For example, the dropout level in each layer, the batch sizes etc. Dropout is a technique that involves random removal of arrows from the network during the internal learning steps. It can also be set in the first step by specifying a dropout layer. Finally, we define a loss function such as $l_2$ loss for regression or cross-entropy for classification and fit the network parameters by optimizing it using SGD techniques \cite{LeCun2015,Goodfellow2016}. The backpropagation algorithm is an elegant way to exploit the layered structure of the network to compute gradients efficiently. It works by transforming the learning process on all layers into SGD runs. There are other techniques in deep learning such as asynchronous SGD (AsyncSGD) that can be distributed and better exploit systems with more resources \cite{Dean2012}. 

As a proof of concept we illustrate why deep learning tasks fit our framework we review recent advances in AsyncSGD \cite{45187,Wei2016}. These algorithms work by having two types of machines in the system: some are workers tasked with optimization and some are listeners of the workers that are tasked with updating the solutions and sending updated parameters and intermediate solutions to the workers. The goal of developing fast asynchronous optimization techniques for machine learning is clear: practitioners will be able to train better models and do it either faster or with different distributed systems. However, these algorithm aggravate the uncertainty aspect of the process, mainly because of \textit{staleness}: when performing distributed AsyncSGD we may have a worker that optimizes a set of parameters that are too old (i.e., were already updated substantially by some other worker). For example, we do not know in advance what is the best partition of our machines into workers vs. listeners.

Specifically \cite{45187} tested different variants of AsyncSGD including a semi synchronous alternative called Sync-OPT. The authors illustrated several properties, on different datasets including ImageNet and MNIST, that are relevant to us on different datasets. First, the experiments nicely show that by using AsyncSGD an almost linear improvement in running time can be achieved as a function of the number of used GPUs. However, using too many worker GPUs per task (>90 as a rule of thumb, can result in >5 hours slow down) may be too disruptive and the algorithm may become inefficient. Second, tens and even a hundred GPUs can be exploited to solve problems. Third, different variants even when using the same computational resources may differ substantially in precision (even up to 4-6\%). 

\subsubsection{Mixed Integer Linear Programming}

This is general framework that can be used for many pattern recognition problems, see \cite{BenBachouch2012} for example. Rough estimate of the importance of this field quickly covers thousands of commercial companies from more than forty different fields that use the framework \cite{gurobiweb}. An MILP program is a problem with: (1) a linear objective function $f^T x$, where f is a column vector of constants and x are variables, (2) bounds and linear constraints, non-linear constraints are not allowed, and (3) restrictions on some of the variables to be integers \cite{Wolsey2007}. A standard representation of the problem is typically written:

\begin{equation*}
\begin{aligned}
& \underset{x}{\text{minimize}}
& & f^T x \\
& \text{subject to}
&  & Ax \le b\\
& & & A_{eq} x = b_{eq}\\ 
& & & l_b \le x \le u_b \\
& & & \forall j\in S, S \subseteq \{1,…,|x|\} \ \   x_j \in \mathbb{Z}
\end{aligned}
\end{equation*}

Even though MILP is NP-Complete (easily via a reduction from clique) many solvers have been presented previously that work well in many practical cases even when the number of variables and constraints reach tens of thousands. There are both commercial and open source available solvers. Most notable commercial tools are CPLEX (IBM) \cite{CPLEX2009}, Gurobi \cite{Lee2007}, and LINDO \cite{Lin2009}. Even Microsoft Excel and MATLAB support MILP solving. Notable open source tools are CLP (from COIN-OR \cite{Martin2010}) and glpk. There are more than a dozen other available tools. Some of these tools use a branch-and-cut algorithm to guide the search space. These algorithms are highly dependent on the order of the variables, which can be used as a randomness source for the competition. 

\subsection{Data size example: small to medium size problem}
Assume we deal with a small to medium size problem that is at most 100MB of data and the solution size is ~1\% of the problem size which is less than 1MB. For example, finding the maximal clique: the input may be a matrix of $n^2$ for dense graphs, while the output in $n$. Therefore, for any $n>100$ the output is less than 1\%. Assuming average upload rate of a user is $8Mb \rightarrow 1MB$. In this case the client uploads the problem in a couple of minutes. Assuming average download rate of $24Mb \rightarrow 3MB$. Therefore, the miner downloads the problem in roughly half a minute. As long as solving the problem takes more than half a minute, the miner can be kept busy (http://www.speedtest.net/reports/united-states/). Solving the maximum clique problem exhaustively will take $2^n$. A farm with 100 GPUs can do $800T$ flops.

\bibliographystyle{ieeetr}

\appendix

\section{Appendix A: Abbreviations}
\begin{itemize}
\item dTMN: Dynamic Task masternode network
\item CT: Computational task
\item GD: Gradient Descent
\item MILP: Mixed Integer Linear Programming
\item PoAW: Proof-of-Accumulated-Work
\item PoD: Proof-of-Deposit
\item PoW: Proof-of-Work
\item PoS: Proof-of-Stake
\item PoSe: Proof-of-Service
\item PoSt: Proof-of-Spacetime
\item PoRep: Proof-of-Replication
\item PF: Promised fee
\item SGD: Stochastic Gradient Descent
\item SSA: Suboptimal solution adversary
\end{itemize}

\section{Appendix B: Parameters}
\begin{itemize}
	\item $r_s$: minimal number of replicates in a TMN
	\item $r>1$: the expected profit factor of a PoS miner
	\item $fee_{solve}$: the total amount that the client is willing to pay for a task $CT$
	\item $fee_{tr}$: a standard transaction fee that is suggested in order to add the transaction into the network
	\item $p_{vstake}$: the percentage of earned useful work amount that is created as virtual stakes
	\item $p_{pool1}, p_{pool2}$
	\item $p_{pools}=p_{pool1}+p_{pool2}$
	\item $B_{dist}$: the time window for running pool distribution fee algorithms
\end{itemize}

\section{\textbf{Minimum distance in graphs}} Let $G=<V,E>$ be an undirected graph with non-negative edge weights, and a pair of nodes $u,v\in V$. The distance between $u$ and $v$ is the minimal number of steps required to traverse the graph such that we start from $v$ and reach $u$. This task does not satisfy resistance, hiding, and puzzle friendliness. In addition, validating the distance is not exponentially easier than solving the problem.
\pf{
\begin{enumerate}
	\item It is straightforward to find input pairs with the same output: we can easily change $G=<V,E>$ into a new input $G'=<V \cup \{ x \} , E>$ and the output for u,v is the same. We simply added a new node $x$ that has no edges going in or out. These two instances have the exact same solution.
	\item This function is not hiding: knowing the minimal distance between $u$ and $v$, by definition, gives us information about the underlying graph structure.
	\item Puzzle friendliness is also violated because: (1) finding the minimal distance between $u$ and $v$ is solvable in $O(|E|+|V|  log⁡|V|)$, and (2) there are input instances for which there are better algorithms than the worst-case: if the underlying graph is a tree, then there is an algorithm that solves the problem in $O(|V|)$. 
	\item Cheap validation: validation can be as hard as solving the original problem because it requires (in the worst-case) resolving the problem. In contrast, for hash functions, validation is exponentially cheaper than solving the puzzle. \qed
\end{enumerate}
}

\section{Storage blockchains}

\subsubsection{Filecoin}
Filecoin manages two markets. First, the \textit{Storage Market} allows clients to pay the storage miners for storing data. This market audits \textit{in-chain}: the management process is done using transactions that are recorded in the blockchain. Second, the \textit{Retrieval Market} allows clients to retrieve data by paying data retrieval miners for delivering data. This market audits off-chain: there is a private process taken between the client and the provider and the internal steps are not published in the blockchain. 
In both markets, clients submit \textit{bid} orders to the market to request a service. In response, the miners submit ask orders to offer a service, when both parties agree on a price, they jointly create a deal order. In both cases the output deal is signed digitally and is published as a transaction in the blockchain. The off-chain component of data retrieval is that the client-miner negotiation is done privately by them.
A filecoin block is divided into three parts: 
\begin{enumerate}[I.]
\item Orderbook: a data structure that holds in-chain bid, ask, and deal transactions of the Storage Market.
\item Transactions: records that hold payment transactions and client transaction. Filecoin’s transactions act as collateral - clients deposit the funds specified in the order, guaranteeing commitment and availability of funds during settlement. 
\item Allocation table: holds pointers to the stored data. The miners’ storage is partitioned into sectors, where each sector contains data pieces assigned to the miner. The Network keeps track of each sector assignments to miners via this allocation table. When a storage miner sector is filled, the sector is sealed. Sealing is a slow, sequential operation that transforms the data in a sector into a replica, a unique physical copy of the data that is associated with the public key of the storage miner.
\end{enumerate}

Trustless storage systems have to prove that it stores the data. This is achieved in Filecoin using a method called Proof-of-Storage. The miner publishes a proof on chain using a challenge-response protocol every $t$ time. Filecoin introduced new methods that implement this protocol such that a miner can prove the data exists in storage. Their challenge-response mechanism is called PoRep (proof of replica) and PoSt (proof of spacetime) and it is different from other blockchains like Storj. PoSt schemes allow a user to check if a storage provider is storing the outsourced data for a certain time range (and not only at the time of the challenge).

In order to retrieve the data efficiently, the storage system cannot wait for auditing each operation in-chain. Therefor filecoin uses an incremental micropayment scheme. In this scheme, two party setup a multi signature \textit{channel}, funds are placed into this channel. This channel is represented as an entry on the public ledger. In order to spend funds from the channel, both parties must agree on the new balance. The current balance is stored as the most recent transaction signed by both parties. Bitcoin lightning network is an example for a system that uses such schemes \cite{Poon2016}.

Data encryption: The encryption is made only by the client. There is no additional mechanism.

Blockchain maintenance: Filecoin is based on a Proof-of-Stake consensus protocol where the stakes are based on the storage assigned to each miner (see section 6.2.3 in the paper). This is an elegant PoS solution with is similar in spirit to our vstake mechanism. However, Filecoin does not address the N@S problem, nor is it designed to handle computational work. \\

Flow in the Storage Market:
\begin{itemize}
 \item To store the data, clients first publish a bid transaction on the blockchain. Miners can then see it and in response publish their ask transaction. If the negotiation is successful, the client and the miner both sign a deal and publish it on the blockchain.
 \item A pointer to the data is written in the allocation table.
 \item The miner is paid for storing the data over time. Every t time it publishes a PoSt and gets paid a fraction per proof.
 \end{itemize}

Flow in the Retrieval Market:
\begin{itemize}
\item Retrieving the data is done off-chain through a negotiation between the client and the miners holding the data (exchange ask, bid, deal p2p without auditing it on-chain). When the deal is done and signed by both parties, the client receives the data. 
\item The payment is done using a micropayment mechanism, which is similar to a lightning network, and a payout signature is published when retrieval is done (for payment extraction).  Thus, the public ledger contains information that the deal was completed.
\end{itemize}

\subsubsection{Storj}

Storj is an application running on the Ethereum blockchain. Storj uses an ERC-20 token called STORJ for making payments.  Storj uses a trusted server called bridge. The bridge is used for delegating trust to dedicated server that manages data ownership and payments. 

The Storj market is based on a contract negotiation between the storage provider and the client. If they both agree on a price then data can be stored. The client splits the file into shards (encrypted parts of the file). The negotiation is mediated by the bridge. The bridge can sniff the network and find storage providers. Note that Storj heavily relies on bridges throughout the pipeline and it is therefore not fully decentralized.

After the negotiation, the client sends the shards to the miners. Payment is done by micropayment mechanism (like in lightning networks). The bridge verifies that the miner holds the shard periodically by Proof-of-Replication mechanism (challenge-response) and pays the storage provider on every proof. 

Retrieving the data is performed by negotiation of the bridge and the miners holding the data. Payment is also carried by a micropayment mechanism. 

Data encryption: Encryption is made by the user. The bridge also takes a park in maintaining the privacy of the data by dividing the file into shards. Only the client and the bridge know where these data are. As a result, gathering all the data for going back to the information in the input files cannot be done by a malicious adversary.


\end{multicols}
\end{document}